
\magnification=\magstep 1
\overfullrule=0pt
\hfuzz=16pt
\voffset=0.0 true in
\vsize=8.8 true in
\baselineskip 20pt
\parskip 6pt
\hoffset=0.1 true in
\hsize=6.3 true in
\nopagenumbers
\pageno=1
\footline={\hfil -- {\folio} -- \hfil}

\ \hfill (revised August 2000)

\ 

\ 

\centerline{\bf Nuclear Spin Qubit Dephasing Time in the}

\centerline{\bf Integer Quantum Hall Effect Regime}

\vskip 0.32in

\centerline{\bf Dima Mozyrsky{\rm ,} Vladimir Privman {\rm and} Israel D.~Vagner$^*$}

\vskip 0.24in

\noindent{Department of Physics, Clarkson University,
Potsdam, New York 13699--5820, USA}

\noindent{$^*$also at Grenoble High Magnetic Field Lab,
Max-Planck-Institut f\"ur Festk\"orperforschung, and
Centre National de la Recherche Scientifique,
BP 166, Grenoble 9, F-38042, France}

\vskip 0.32in

\centerline{\bf ABSTRACT}

We report the first theoretical estimate of the nuclear-spin dephasing time $T_2$ owing to the spin interaction with the two-dimensional electron gas, when the latter is in the integer quantum Hall state, in a two-dimensional heterojunction or quantum well at low temperature and in large applied magnetic field. We establish that the leading mechanism of dephasing is due to the impurity potentials that influence the dynamics of the spin via virtual magnetic spin-exciton scattering. Implications of our results for implementation of nuclear spins as quantum bits (qubits) for quantum computing are discussed.

\vfill


\vfil\eject

\noindent {\bf 1.\ Introduction}

Recent ideas [1-3] of utilizing nuclear spins in semiconductor quantum wells and heterojunctions as quantum bits (qubits) for quantum computation have generated new emphases in the studies of nuclear-spin relaxation and, especially, quantum decoherence, in such systems. In this work we consider the case of the integer $\nu=1$ quantum-Hall state [4]. The two-dimensional electron gas is then in a nondissipative state. Since the electrons mediate the dominant interaction [1,5,6] between nuclear spins, it is reasonable to expect that relaxation times of the latter, as well as decoherence/dephasing effects, will occur on large time scales.

Solid-state proposals for quantum computation [1-3] with nuclear spins are all presently theoretical. Related proposals to {\rm utilize} quantum dots [7-15] are also at present all in the theory stage. Usually, non-zero-nuclear-spin atoms will be considered placed [1] by modern ``atomic engineering'' techniques in a host material of zero nuclear spin isotope. In order to allow positioning with respect to other features of the system, such as gate electrodes [2], and making replicas [1], etc., the nuclear-spin separation will be larger than the atomic size, typically, of order 20 to 100\AA. At these separations, the direct magnetic dipole-dipole interaction of the nuclear spins is negligible.

The dynamics of the nuclear spins is governed by their interactions with each other and with their environment. In the regime of interest, these interactions are mediated by the two-dimensional electron gas.
Various time scales are associated with this dynamics. The relaxation time $T_1$ is related to energy exchange and thermalization of the spins. Quantum mechanical decoherence and dephasing will occur on the time scale $T_2$. {\rm The latter processes correspond to demolition of quantum-mechanical superposition of states and erasure of phase information, due to interactions with the environment.} Generally, there are many dynamical processes in the system, so the times $T_1$ and $T_2$ may not be uniquely, separately defined [16,17]. Theoretically and experimentally, it has been established that processes of energy exchange are slow at low temperatures, so $T_1$ is very large, but there still might be some decoherence owing to quantum fluctuations. Generally, for various systems, there are extreme examples of theoretical prediction, ranging from no decoherence to finite decoherence [18,19,20] at zero temperature, depending on the model assumptions.

In order to consider control (``programming'') of a quantum computer, we have to identify the time scale $T_{\rm ext}$ of the single-spin rotations owing to the interactions with an external NMR magnetic field. We also identify the time scale $T_{\rm int}$ associated with evolution owing to the pairwise spin-spin interactions. The preferred relation of the time scales is $ T_1,\,  T_2 \, \gg \, T_{\rm ext},\, T_{\rm int} \,$, which is obviously required for coherent quantum-mechanical dynamics. 

The aim of this work has been to advance theoretical understanding of the time-scales of interest for the quantum computer proposal [1] based on nuclear spins in a two-dimensional electron gas, with the latter in the integer quantum Hall effect state obtained at low temperatures, of order 1K, and in high magnetic fields, of several Tesla, in two-dimensional semiconductor structures [4]. This system is a promising candidate for quantum computing because the nuclear spin relaxation time $T_1$ can be as large as $10^3\,$sec. In the summarizing discussion, Section~5, we discuss and compare the values of all the relevant time scales.

Our main result, presented in Sections~2 through 4, is {\it the first theoretical calculation\/} of the nuclear-spin dephasing/decoherence time scale $T_2$ for such systems. We note that the recent study [21-23] of the nuclear-spin relaxation time $T_1$, 
has relied heavily on the accepted theoretical and experimental views of the properties and behavior of the {\it electronic state\/} of the two-dimensional electron gas in the quantum Hall regime. These electronic properties have been a subject of several studies [4-6,21-29]. We  
utilize these results in our calculation as well.

{\vskip 0.16 true in}\noindent {\bf 2.\ The Model}

We consider a single nuclear spin coupled to a two-dimensional electron gas in a strong magnetic field, $B$, along the $z$ axis which is perpendicular to the two-dimensional structure. Assuming nuclear spin-${1\over 2}$, for simplicity, we write the Hamiltonian as 

$$H = H_n + H_e + H_{ne} + H_{\rm imp} \eqno (1)$$

\noindent
Here the first term is the nuclear spin interaction with the external magnetic field, $H_n = -{1\over 2}\gamma_n B \sigma_z$, 
where $\gamma_n$ includes $\hbar$ and the nuclear $g$-factor, and $\sigma_z$ is a Pauli matrix.

The second term is the electronic component of the total Hamiltonian (1). Within the free-electron nomenclature, the Fermi level lies in between the two Zeeman sub-levels of the lowest Landau level. The spin-up sub-level is then completely occupied, so the filling factor is $\nu = 1$, while the spin-down sub-level is completely empty; note that the relevant effective electronic $g$-factor in typically negative. In fact, the calculation need not be limited to the lowest Landau level. Here, however, to avoid unilluminating mathematical complications, we restrict our attention to the lowest level, as has been uniformly done in the literature [24-27].

The last two terms in (1) correspond to the nuclear-spin electron interactions and to the effects of impurities. These will be addressed shortly. The magnetic sub-levels are actually broadened by impurities. 
At low temperatures, the $\nu=1$ system is in the quantum Hall state. The interactions of the two-dimensional electron gas with the underlying material are not shown in (1). They are accounted for phenomenologically, as described later. 

The electron-electron interactions are treated within an approximate quasiparticle theory which only retains transition amplitudes between Zeeman sub-levels. The elementary excitations of the electron gas are then well described as magnetic spin excitons, or spin waves, [24-27]. The spin excitons are quasiparticles arising as a result of the interplay between the Coulomb repulsion of the electrons and their exchange interaction. A creation operator of a spin exciton with a two dimensional wave vector ${\bf k}$ can be written in terms of the electronic creation operators $a^{\dag}$ in the spin-down Zeeman sub-level and annihilation operators $b$ in the spin-up sub-level as

$$A_{\bf k}^{\dag} = \sqrt{2\pi \ell^2 \over L_x L_y} \sum_p e^{i\ell^2 k_x p}  a^{\dag}_{p+{k_y \over 2}} b_{p-{k_y \over 2}}\eqno (2)$$

\noindent
Here $\ell=\sqrt{c\hbar/eB}$ is the magnetic length, and the $p$-summation is taken in such a way that the wave number subscripts are quantized in multiples of $2\pi /L_y$. Note that expression (2) assumes the Landau gauge, which is not symmetric under $x \leftrightarrow y$. {\rm For our purposes, the following expression for the dispersion relation of the excitons [24,25] provides an adequate approximation,}

$$E_{\bf k} = \Delta + \left({\pi \over 2}\right)^{1/2} \left({e^2 \over \epsilon \ell}\right)
{\ell^2 k^2\over 2} \eqno (3)$$           

\noindent
Here $\Delta = |g|\mu_B B$, where $\mu_B$ is the Bohr magneton, and $g$ is the electronic $g$-factor, and $\epsilon$ is the dielectric constant of the material. It has been pointed out [6,23] that the gap $\Delta$ in the excitonic spectrum suppresses nuclear spin relaxation at low temperatures.
The electronic Hamiltonian can be written in terms of the spin exciton operators as 

$$H_e = {\cal E}_0 + \sum_{\bf k} E_{\bf k} A_{\bf k}^{\dag} A_{\bf k} \eqno (4)$$

\noindent
where the $c$-number ${\cal E}_0$ is the spin-independent ground state energy of the electron gas. This description of the electronic gas is appropriate only for low density of excitons, which is the case in our calculation, as will be seen later. 

We now turn to the third term in (1), the interaction between the electrons and nuclear spins. It can be adequately approximated by the hyperfine Fermi contact term

$$H_{ne} = {8\pi \over 3} \gamma_n g \mu_B {\bf I}_n \cdot \sum_e {\bf S}_e 
\delta^{(3)} \left({\bf r}_e - {\bf R}_n\right) \eqno (5)$$  

\noindent
Here $\hbar {\bf I}_n$ and $\hbar {\bf S}_e$ are nuclear and electronic spin operators, respectively, and ${\bf r}_e$ are the electron coordinates. The nuclear coordinate ${\bf R}_n$ can be put equal to zero. Such an interaction can be split into two parts 

$$H_{ne} = H_{\rm diag} + H_{\rm offdiag} \eqno (6)$$

\noindent
where $H_{\rm diag}$ corresponds to the coupling of the electrons to the diagonal part of nuclear spin operator ${\bf I}_n$, and $H_{\rm offdiag}$ --- to its off-diagonal part.

The diagonal and off-diagonal contributions can be rewritten in terms of electronic creation and annihilation operators as

$$H_{\rm diag} = {(8\pi/3)\gamma_n g \mu_B |w_0 (0 )|^2 \over \sqrt{\pi} L_y \ell d} \sum_{k,q} e^{-{\ell^2 \over 2} \left(k^2 + q^{2}\right)} \sigma_z \left( a_k^{\dag} a_{q} - b_k^{\dag} b_{q}\right) \eqno (7) $$

$$H_{\rm offdiag} = {(8\pi/3)\gamma_n g \mu_B |w_0 (0 )|^2 \over \sqrt{\pi} L_y \ell d} \sum_{k,q} e^{-{\ell^2 \over 2} \left(k^2 + q^{2}\right)} \left( \sigma^+ b_k^{\dag} a_{q} + \sigma^-a_k^{\dag} b_{q}\right) \eqno (8) $$

\noindent
Here $\sigma^{\pm} = {1\over 2}\left(\sigma_x \pm i \sigma_y \right)$. The interactions of the electrons of the two-dimensional gas with the underlying material are incorporated phenomenologically through the dielectric constant and $g$-factor, see (3), 
et seq., and 
via $|w_0 (0 )|^2$ and $d$ in (8) above. The latter is the transverse dimension of the effectively two-dimensional region (heterojunction, quantum well) in which the electrons are confined. The quantity $w_0 (0 )$ represents phenomenologically the enhancement of the amplitude of the electron wave function at the nuclear position owing to the effective potential it experiences as it moves in the solid-state material. It is loosely related [5,6] to the zero-momentum lattice Bloch wavefunction at the origin. 

For the purposes of the calculations performed here, with the relevant states being the ground state and the single-exciton states of the electron gas, one can show that the terms in (7) that correspond to different $k$ and $q$ do not contribute, while the remaining sum over $k$ becomes a 
$c$-number, representing the Knight shift of the polarized electrons. 
Thus $H_{\rm diag}$ can be incorporated into the nuclear-spin energy splitting, redefining the Hamiltonian of the nuclear spin as $H_n = {1\over 2}\Gamma\sigma_z$, where
$\Gamma = \gamma_n \left(B+B_{\rm Knight}\right)$. Note that the Knight shift can be used to estimate the value of the phenomenological parameter
$|w_0 (0 )|$ from experimental data.
The off-diagonal coupling (8) can be expressed on terms of the excitonic operators (2) as follows [5,6,23],

$$H_{\rm offdiag} =  {C \over \sqrt{L_xL_y} } \sum_{\bf k} e^{-\ell^2 k^2/4}\left(A_{\bf k}^{\dag}\sigma^- + A_{\bf k}\sigma^+ \right) \eqno (9)$$

\noindent where

$$C= {(8\pi/3)\gamma_n g \mu_B |w_0 (0 )|^2 \over \sqrt{2\pi}\ell d}\eqno(10) $$

\noindent The summations over $k_x$ and $k_y$ are taken over all the integer multiples of $2\pi / L_x$ and $2 \pi / L_y$, respectively.

The last term in (1) describes the interaction of the electrons with impurities and plays a crucial role in nuclear relaxation in the systems of interest. This interaction can written in the spin-exciton representation as [23,26]
 
$$H_{\rm imp} = \left(2i / L_x L_y\right) \sum_{\bf k,q} U\left({\bf q}\right)
\sin\left[\ell^2 \left(k_xq_y-k_yq_x\right)/2\right] A_{\bf k}^{\dag} A_{\bf k + q} \eqno (11)$$ 

\noindent
where $U\left({\bf q}\right)=\int U_{\rm imp}\left({\bf r}\right) e^{i{\bf q}\cdot{\bf r}} d^2 {\bf r}$ is the Fourier component of the impurity potential for electrons in the two-dimensional plane. We will assume [23,26] that the impurity potential has zero average and can be modeled by the Gaussian white noise completely described by its correlator, 
$\langle U_{\rm imp}\left({\bf r}\right) U_{\rm imp}\left({\bf r^{\prime}}\right)\rangle = Q \delta^{(2)} \left({\bf r - r^{\prime}}\right)$.

In summary, the relevant terms in the full Hamiltonian (1) can be expressed solely in terms of the nuclear-spin operators and spin-excitation operators as

$$H = -{1\over 2}\Gamma \sigma_z + \sum_{\bf k} E_{\bf k} A_{\bf k}^{\dag} A_{\bf k} + \sum_{\bf k} g_{\bf k} \left(A_{\bf k}^{\dag}\sigma^- + A_{\bf k}\sigma^+ \right) + \sum_{\bf k,q}
\phi_{\bf k,q} A_{\bf k}^{\dag} A_{\bf k + q} \eqno (12)$$

\noindent
where the explicit expressions for $E_{\bf k}$, $g_{\bf k}$ and $\phi_{\bf k,q}$ can be read off (3), (9) and (10), respectively, and the quantity $\Gamma$ was introduced in the text preceding Eq.~(9). 

{\vskip 0.16 true in}\noindent {\bf 3.\ Energy Relaxation}

In order to set the stage for the calculation of $T_2$, let us first briefly summarize in this section aspects of the calculation of the nuclear-spin relaxation time $T_1$, along the lines of [22,23]. The dominant mechanism for both processes at low temperatures is the interactions with impurities. Thus, both calculations are effectively zero-temperature, single-spin; these assumptions will be further discussed in Section 5. 

We assume that initially, at time $t=0$, the nuclear spin is polarized, while the excitons are in the ground state,
$\;|\Psi \left( 0 \right) \rangle = |-\rangle \otimes |{\bf 0}\rangle $,\ 
where $|-\rangle$ is the polarized-down (excited) state of the nuclear spin and $|{\bf 0}\rangle$ is the ground state of spin-excitons. Since the Hamiltonian (12) conserves the $z$-component of the total spin in the system, the most general wavefunction evolving from $|\Psi \left( 0 \right) \rangle$ can be written as 

$$|\Psi \left( t \right) \rangle = \alpha \left( t \right) |-\rangle \otimes |{\bf 0}\rangle +
\sum_{\bf k} \beta_{\bf k} \left( t \right) |+\rangle \otimes |{\bf 1_k}\rangle \eqno (13)$$

\noindent
with $|+\rangle$ corresponding to the nuclear spin in the ground state and $|{\bf 1_k}\rangle$ describing the single-exciton state with the wave vector ${\bf k}$. Equations of motion for the coefficients $\alpha$ and $\beta_{\bf k}$ can be easily derived from the Schr\"odinger equation:

$$i\hbar \dot\alpha = {1 \over 2}\Gamma \alpha + \sum_{\bf k} g_{\bf k} \beta_{\bf k} \eqno (14)$$

$$i\hbar {\dot\beta_{\bf k}} = -{1 \over 2}\Gamma \beta_{\bf k} +  E_{\bf k} \beta_{\bf k} + 
\sum_{\bf q} \phi_{\bf k,q} \beta_{\bf q} + g_{\bf k} \alpha \eqno (15)$$

\noindent
In order to solve the system of equations (14)-(15), we introduce Laplace transforms, $\tilde{f}\left( S \right)=
\int_0^{\infty}  f ( t  ) e^{-St}dt$, which satisfy

$$iS\hbar \tilde{\alpha} - i\hbar = {1 \over 2}\Gamma \tilde{\alpha} + \sum_{\bf k} g_{\bf k} \tilde{\beta}_{\bf k} \eqno (16)$$

$$iS\hbar \tilde{\beta}_{\bf k} = -{1 \over 2}\Gamma \tilde{\beta}_{\bf k} +  E_{\bf k} \tilde{\beta}_{\bf k} + \sum_{\bf q} \phi_{\bf k,q} \tilde{\beta}_{\bf q} + g_{\bf k} \tilde{\alpha} \eqno (17)$$

Let us first solve (16)-(17) for the case when the interaction of spin-excitons with impurities is switched off, i.e., $\phi_{\bf k,q}=0$. After some algebra we obtain
$$ {1 \over \tilde{\alpha}(s)}=s + {i \over \hbar}\sum_{\bf k} {g_{\bf k}^2 \over is\hbar+\Gamma-E_{\bf k}} \eqno (18)$$

\noindent
where we have shifted the variable: $s=S+i\Gamma/(2\hbar)$, which only introduces an noninteresting phase factor.  

In the absence of the hyperfine interaction, i.e., for $g_{\bf k}=0$, $\tilde \alpha (s)$ in (18) has only the pole at $s=0$. When the interaction is switched on, the pole shifts from zero. This shift can be calculated in a standard way, within the leading order perturbative approach, by taking the limit $s \rightarrow 0$, so that ${1 \over i\hbar s^+ +\Gamma -E_{\bf k}} \rightarrow {\cal P}{1 \over \Gamma -E_{\bf k}}-i\pi \delta\left(\Gamma-E_{\bf k}\right)$, where ${\cal P}$ denotes the principal value. This type of approximation is encountered in quantum optics [30]. 
The relaxation rate and the added phase shift of the nuclear-spin excited-state probability amplitude $\alpha(t)$ are given by the real and imaginary parts of the pole,
${1\over T_1} = {2\pi \over \hbar}\sum_{\bf k} g_{\bf k}^2 \delta\left(\Gamma-E_{\bf k}\right)$
and $\omega = {\cal P} \sum_{\bf k} {g_{\bf k}^2 \over \Gamma-E_{\bf k}}$ respectively,
so that $\alpha (t) \propto e^{-t/(2T_1) + i \omega t}$. 
It is obvious that due to the large gap in the spin-exciton spectrum (3), $\Gamma \ll \Delta$, the energy conservation required by the delta function above can never be satisfied, and so in the absence of interaction with impurities, $T_1=\infty$. It also transpires that $T_2$ is infinite [30], as will become apparent later. 

Interactions with impurities, described by the last term in (12), will modify the solution of (16)-(17), and, as a consequence, the energy conservation condition. In particular, if the impurity potential is strong enough, it can provide additional energy to spin-excitons, so that their energy can fluctuate on the scale of order $\Gamma$ thus making nuclear-spin relaxation possible. This mechanism [22,23] corresponds to large fluctuations of the impurity potential $U ({\bf r})$, which usually occur with a rather small probability, so $T_1$ is very large for such systems. 

In order to carry out the above program quantitatively, one has to solve the system of equations (16)-(17) with nonzero $\phi_{\bf k,q}$. Such a solution is only possible within an approximation. One can introduce the effective spin-exciton self-energy $\Sigma_{\bf k}$ in (18),
so that $ {1 \over i\hbar s+\Gamma-E_{\bf k}} \to {1 \over i\hbar s+\Gamma-E_{\bf k} + \Sigma_{\bf k}}$.
An integral equation for $\Sigma_{\bf k}$ can then be derived, 
taking the continuum limit in (16)-(17). Solving this equation 
would allow one to calculate the relaxation rate from (18). However, in order to satisfy the energy conservation, we 
require $\Gamma-E_{\bf k} + \Sigma_{\bf k}=0$, so the self-energy 
should be rather large, of order $E_{\bf k}$. Therefore, as a 
result of the spectral gap of the excitons, the perturbative 
approach is inadequate as it automatically assumes that 
$|\Sigma_{\bf k}| \ll |E_{\bf k}|$. Instead, a certain 
variational approach [23,31] has been adapted to evaluate $T_1$, consistent with the experimental values [33,34] of order 10$^3\,$sec; for further discussed see Section 5.

{\vskip 0.16 true in}\noindent {\bf 4.\ Dephasing Mechanism}

We argue that in order to calculate the phase shift due 
to the impurity potential, one can indeed use the perturbative 
solution of (16)-(17). Indeed,  phase shifts result in virtual 
processes that do not require energy conservation and 
therefore are dominated by relatively small fluctuations 
of the impurity potential simply because large fluctuations are very rare.   
Moreover, the terms in the sum in (18) that contribute to the relaxation rate, 
do not contribute to the phase shift, see the discussion above. This consideration 
also applies when the self-energy is introduced.

{\rm One can show that the contribution to dephasing linear in $\phi_{\bf k,p}$ vanishes due to symmetry. Thus, let us solve (16)-(17) perturbatively up to the second order in $\phi_{\bf k,p}$ and perform the inverse Laplace transform of $\tilde \alpha (s)$. Within this approximation, the pole of $\tilde \alpha (s)$ in the complex-$s$ plane is imaginary, so that $|\alpha(t)|=1$. 
We conclude that $\alpha(t)\propto e^{i\omega_U t}$ and 
$\beta_{\bf k} ( t ) =0$, where the part of the phase-shift responsible for dephasing is}

$$\omega_U = - {1 \over \hbar} \sum_{\bf k} {g_{\bf k} \over E_{\bf k}}
\sum_{\bf q} {\phi_{\bf k,q} \over E_{\bf k-q}} \sum_{\bf p}
{\phi_{\bf k-q,p} g_{\bf k-q-p}\over E_{\bf k-q-p}} \eqno (19)$$

\noindent {\rm The zeroth-order term in (19) was dropped  as irrelevant for our calculation of the dephasing time. Since $\Gamma$ is much smaller than $E_{\bf k}$, it was also omitted.} 

As expected, the perturbative solution does not describe 
the energy relaxation ($T_1$), but it does yield the {\rm additional} phase shift due 
to the impurity potential. We will see shortly that this phase shift, 
when averaged over configurations of the impurity potential, produces 
a finite dephasing time, $T_2$. 

Let us consider the reduced density matrix of the nuclear spin, given by

$$ \rho_n ( t ) = \big[\,{\rm Tr}_e |\Psi \left( t \right) \rangle\langle 
\Psi \left( t \right)| \,\big]_U \eqno (20)$$ 

\noindent
recall (13). Here the trace is partial, taken over the states of the 
spin-excitons, while the outer brackets denote averaging over the 
impurity potential. The trace over the spin-excitons can be carried 
out straightforwardly because within the leading-order perturbative 
approximation used here they remain in the ground state; all 
excitations are virtual and contribute only to the phase shift. 
The diagonal elements of $ \rho_n ( t )$ are not influenced by 
virtual excitations and remain constant. 

The off-diagonal elements of $ \rho_n ( t )$ contain the factors 
$e^{\pm i \omega_U t}$. It is the averaging of these quantities over the 
white-noise impurity potential $U({\bf r})$ that yields dephasing 
of the nuclear spin. 
{\rm In order to proceed, let us rewrite (19) more explicitly. From (9)-(11), after changing summation index ($\bf k \to k-{q+p \over \rm 2}$) in the first sum in (19) we obtain}

$$\omega_U = {4 C^2 \over \hbar (L_x L_y)^3} \sum_{\bf q,p}U({\bf q})U({\bf p})
e^{-{\ell^2 \over 8}\left({\bf p+q}\right)^2} \sum_{\bf k} {e^{-{\ell^2 \over 2}{\bf k}^2}
\sin{{\ell^2 \over 2}\left[{\bf k+{p \over \rm 2},q}\right]_z}\sin{{\ell^2 \over \rm 2}\left[{\bf k-{q \over \rm 2},p}\right]_z} \over E_{\bf k+{q+p \over \rm 2}}E_{\bf k+{q-p \over \rm 2}}E_{\bf k-{q+p \over \rm 2}}}
\eqno (21)$$

\noindent
Here we use the following shorthand notation for the $z$ component of a vector product $\left[{\bf k,q}\right]_z = k_x q_y - k_y q_x$.

It is appropriate to assume that impurity potentials are short-range, i.e., $a \ll \ell$, where $a$ is the scale of variation
of $U_{\rm imp}({\bf r})$. This assumption and the white-noise property of the impurity potentials, are required to make the problem amenable to analytical calculation. Thus, the main contribution to the Fourier transform $U({\bf p})$, dominating the summation in (21), comes from large wavevectors ${\bf p}$ (and ${\bf q}$), of order $a^{-1} \gg \ell^{-1}$. 
Therefore one can replace the exponent $e^{-{\ell^2 \over 8}\left({\bf p+q}\right)^2}$ by the Kronecker symbol $\delta_{\bf q,-p}$, to obtain a simplified expression for $\omega_U$

$$\omega_U = {4 C^2 \over \hbar (L_x L_y)^3} \sum_{\bf p}U({\bf p})U({\bf -p})
\sum_{\bf k} {e^{-{\ell^2 \over 2}{\bf k}^2}
\sin^2{{\ell^2 \over 2}\left[{\bf k,p}\right]_z} \over E^2_{\bf k}E_{\bf k+p}}
\eqno (22)$$

\noindent
Now the sum over ${\bf k}$ can be carried out because for ${\bf p} \gg {\bf k}$, we can assume that $E_{\bf k+p} \simeq E_{\bf p} \simeq E_c (\ell^2 / 2){\bf p}^2 $, 
where $E_c = \left({\pi / 2}\right)^{1/2} \left[{e^2 / (\epsilon \ell)}\right]$.

Moreover for large ${\bf p}$, the factor $\sin^2 \left\{ (\ell^2 / 2)\left[{\bf k,p}\right]_z\right\}$ can be replaced by its average, ${1 / 2}$. Finally, we get 

$$ {1 \over L_x L_y} \sum_{\bf k} {e^{-{\ell^2 \over 8}{\bf k}^2} \sin^2{{\ell^2 \over 2}\left[{\bf k,p}\right]_z} \over E^2_{\bf k}E_{\bf k+p}} \simeq {1 \over E_c \ell^2 {\bf p}^2} {1 \over 2\pi} \int k dk {e^{-{\ell^2 \over 8}k^2} \over \left(\Delta + E_c {\ell^2 \over 2} k^2\right)^2} \eqno(23)$$

\noindent
The integral can be evaluated explicitly; specifically, for ${\Delta \over E_c} \ll 1$ we get

$$ \int_0^{\infty} k dk {e^{-{\ell^2 \over 8}k^2} \over \left(\Delta + E_c {\ell^2 \over 2} k^2\right)^2} \simeq {1 \over \ell^2 E_c \Delta} \eqno(24)$$

\noindent
so that 

$$\omega_U = {2 C^2 \over \hbar \pi E_c^2 \ell^4 \Delta}{1 \over (L_x L_y)^2} \sum_{\bf p}{U({\bf p})U({\bf -p})\over {\bf p}^2}\eqno (25)$$

Recall that we have assumed the white-noise distribution for the impurity potential, $\langle U_{\rm imp}\left({\bf r}\right) U_{\rm imp}\left({\bf r^{\prime}}\right)\rangle = Q \delta^{(2)} \left({\bf r - r^{\prime}}\right)$. This corresponds to the following probability distribution functional for the Fourier-transformed potential,  

$$P\left[U(p)\right] = N \exp{\left[-{1 \over 2Q L_x L_y} \sum_{\bf p} U({\bf p})U({\bf -p})
\right]} \eqno (26)$$  

\noindent
The latter expression, and other approximations assumed earlier, allow to reduce the averaging of $e^{i\omega_Ut}$ to a product of Gaussian integrations. The off-diagonal elements of the nuclear-pin density matrix are, thus,

$$\rho_{01} \sim \prod_{\bf p}\left( 1 - {i\tau \over L_x L_y {\bf p}^2}\right)^{-{1\over 2}}
=\exp{\left[-{1 \over 2} \sum_{\bf p} \ln{\left(1-{i\tau \over L_x L_y {\bf p}^2}\right)}
\right]} \eqno (27)$$

\noindent
where $\tau = {4 Q C^2 t / ( \hbar \pi E_c^2 \ell^4 \Delta)}$.

We are interested in real part of the sum in (27), which represents de\-co\-he\-rence/de\-phas\-ing of the nuclear spin. The off-diagonal elements decay exponentially as

$$\rho_{01} \sim \exp{\left[-{1 \over 4} \sum_{\bf p} \ln{\left(1+{\tau^2 \over (L_x L_y)^2 {\bf p}^4}\right)}\right]} \eqno (28)$$

\noindent
The summation over ${\bf p}$ in (32) can be converted into integration

$$ \sum_{\bf p} \ln{\left(1+{\tau^2 \over (L_x L_y)^2 {\bf p}^4}\right)} = {L_x L_y \over (2\pi)^2} \int d^2 {\bf p} \ln{\left(1+{\tau^2 \over (L_x L_y)^2 {\bf p}^4}\right)} \eqno(29) $$

\noindent
Explicit calculation then yields the result that $\rho_{01} \sim e^{-{\tau / 16}}$ or $\rho_{01} \sim e^{-{t / T_2}}$, where

$$ T_2 = {2\hbar \ell^2 \Delta \over U_2 C^2} \eqno (30)$$

\noindent
with $ U_2 = { Q / ( 2\pi \ell^2 E_c^2)} $.

{\vskip 0.16 true in}\noindent {\bf 5.\ Results and Discussion}

The quantity $U_2$ characterizes the strength of the impurity potential with respect to the Coulomb interactions [26]. Let us summarize typical parameter values [26] for a GaAs heterojuction, which is the system best studied in the literature. For magnetic field value $B=10 \,$T, we have the following values of parameters,
$\ell=0.8\times 10^{-8}\,$m, \ $E_c = 3\times 10^{-21}\,$J, \ $C = 2.5\times 10^{-36}\,$J$\,$m, \ $\Delta=4.6 \times 10^{-23}\,$J. From experimental data for electronic mobility, one then estimates [26] $U_2 \simeq 0.0025$, yielding
$ T_2 \simeq 40 \,{\rm sec}$. We emphasize that this is an order 
of magnitude estimate only, because of the uncertainty 
in various parameter values assumed and the fact that the parameters, especially the strength of the disorder, may vary significantly from sample to sample. 
For instance, there is another estimate of the disorder strength $Q$ available in the literature [23], obtained by fitting the value of $T_1$ to the experimentally measured $10^3\,$sec [33,34], as cited earlier. This yields an estimate for $T_2$ that is smaller, $ T_2 \simeq 0.5\,$sec. Generally, we expect that with typical-quality samples
$T_2$ may be a fraciton of a second or somewhat larger.

Let us point out that to date there are no direct experimental probes of dephasing by the disorder-dominated mechanism identified here for dilute
nuclear-spin positioning appropriate for quantum computing. Such systems where never engineered.
For those materials whose atoms have nonzero nuclear-spin-isotope nuclei, specifically, for GaAs (spins $3/2$), we are aware only of one experiment [35] where
indirect information on dephasing can be obtained from the linewidth. However, in that case the
dipolar interactions cannot be neglected and likely provide the dominant dephasing mechanism. For quantum computing, the host material will have to be isotope-engineered with zero nuclear spins, e.g., Si [2].

Let us now compare various time scales relevant for quantum 
computing applications. The relaxation time $T_1$ is of order 
$10^3\,$sec [23,33,34]. For the spin-spin interaction time 
scale $T_{\rm int}$, values as short as $10^{-11}\,$sec have 
been proposed [1,5]. These estimates are definitely overly 
optimistic and require further work. Since such calculations
require considerations beyond the single-spin interactions,
they are outside the scope of the present work. 
For $T_{\rm ext}$, modern experiments have used NMR field intensities 
corresponding to the spin-flip times of $10^{-5}\,$sec. 
This can be reduced to $10^{-7}\,$sec, and with substantial 
experimental effort, perhaps even shorter times, the main 
limitation being heating up of the sample by the radiation. 

Thus, the present information on the relevant time scales
does not show violation of the condition $T_1,T_2 \gg 
T_{\rm ext},T_{\rm int}$, stated in the introduction, required
for quantum computing. To firmly establish the feasibility of
quantum computing, reliable theoretical evaluation of $T_{\rm int}$
is needed, as well as experimental realizations of few-qubit
systems engineered with nuclear spins positioned as separations
of order 30 to 100$\,${\AA}.

We also note that typical lab samples, for which the 
parameter values used were estimated, have been prepared 
to observe the quantum-Hall-effect plateaus in the resistance. 
The latter requires a finite density of impurities. However, 
for the quantum-computer applications, a much cleaner sample 
would suffice. Indeed, as suggested by our calculations, $T_2$ 
is mostly due to dephasing owing to virtual spin-exciton scattering 
from impurities. Therefore, the value of $T_2$ can be increased 
by using cleaner samples. 

We acknowledge helpful discussions with
Drs. S.~E.~Barrett, M.~L.~Glasser and R.~Mani. This research was supported by the National Security Agency (NSA) and Advanced Research and Development Activity (ARDA) under Army Research Office (ARO) contract number DAAD$\,$19-99-1-0342.

\vfil\eject

\centerline{\bf REFERENCES}

\ 

{\frenchspacing

\item{1.} V. Privman, I.D. Vagner and G. Kventsel, 
Phys. Lett. {\bf A 239}, 141 (1998).

\item{2.} B.E. Kane, Nature {\bf 393}, 133 (1998).

\item{3.} C.M. Bowden and S.D. Pethel, Laser Phys. {\bf 10}, 35 (2000).

\item{4.} {\it The Quantum Hall Effect}, R.E. Prange 
and S.M. Girvin, editors (Springer-Verlag, New York, 1987).

\item{5.} Yu.A. Bychkov, T. Maniv and I.D. Vagner, Solid State
Commun. {\bf 94}, 61 (1995).

\item{6.} I.D. Vagner and T. Maniv, Physica {\bf B 204}, 141 (1995).

\item{7.} D. Loss and D.P. DiVincenzo, Phys. Rev. {\bf A 57}, 120 (1998).

\item{8.} M.S. Sherwin, A. Imamoglu and T. Montroy, Phys. Rev. {\bf A 60}, 3508 (1999).

\item{9.} A. Imamoglu, D.D. Awschalom, G. Burkard, 
D.P. DiVincenzo, D. Loss, M. Sherwin and A. Small, 
Phys. Rev. Lett. {\bf 83}, 4204 (1999).

\item{10.} R. Vrijen, E. Yablonovitch, K. Wang, H.W. 
Jiang, A. Balandin, V. Roychowdhury, T. Mor and D. 
DiVincenzo, Phys. Rev. {\bf A 62}, 012306 (2000).

\item{11.} T. Tanamoto, Physica {\bf B 272}, 45 (1999).

\item{12.} G.D. Sanders, K.W. Kim and W.C. Holton, Phys. Rev. {\bf A 60}, 4146 (1999).

\item{13.} S. Bandyopadhyay, Phys. Rev. {\bf B 61}, 13813 (2000).

\item{14.} X. Hu and S. Das Sarma, Phys. Rev. {\bf A 61}, 062301 (2000).

\item{15.} N.-J. Wu, M. Kamada, A. Natori and H. 
Yasunaga, {\it Quantum Computer Using Coupled Quantum Dot 
Molecules}, preprint quant-ph/9912036.

\item{16.} K. Blum, {\it Density Matrix Theory and 
Applications\/} (Plenum, New York, 1981).

\item{17.} C.P. Slichter, {\it Principles of Magnetic 
Resonance}, Third Edition (Springer-Verlag, Berlin, 1990). 

\item{18.} D. Mozyrsky and V. Privman, J. Statist. Phys. {\bf 91}, 787 (1998).

\item{19.} {\it Mesoscopic Phenomena in Solids}, Modern 
Problems in Condensed Matter Sciences --- Vol. 30, B.L. 
Altshuler, P.A. Lee and R.A. Webb, editors (Elsevier, Amsterdam, 1991).

\item{20.} A.J. Legget, S. Chakravarty, A.T. Dorsey,
M.P.A. Fisher and W. Zwerger, Rev. Mod. Phys. {\bf 59}, 1
(1987) [Erratum {\it ibid.} {\bf 67}, 725 (1995)].

\item{21.} I.D. Vagner and T. Maniv, Phys. Rev. Lett. {\bf 61},
1400 (1988).

\item{22.} D. Antoniou and A.H. MacDonald, Phys. Rev. {\bf B 43}, 11686 (1991). 

\item{23.} S.V. Iordanskii, S.V. Meshkov and I.D. Vagner,
Phys. Rev. {\bf B 44}, 6554 (1991).

\item{24.} Yu.A. Bychkov, S.V. Iordanskii and G.M. Eliashberg,
JETP Lett. {\bf 33}, 143 (1981).

\item{25.} C. Kallin and B.I. Halperin, 
Phys. Rev. {\bf B 30}, 5655 (1984).

\item{26.} C. Kallin and B.I. Halperin, 
Phys. Rev. {\bf B 31}, 3635 (1985).

\item{27.} {\it Quantum Hall Effect}, A.H. MacDonald, Editor 
(Kluwer Academic Publ., Dordrecht, 1989).

\item{28.} {\it Quantum Hall Effect}, M. Stone, Editor 
(World Scientific, Singapore, 1992).

\item{29.} {\it Perspectives in Quantum Hall Effects: Novel 
Quantum Liquids in Low-Dimensional Semiconductor Structures}, 
S. Das Sarma and A. Pinczuk,
Editors (Wiley, New York, 1996). 

\item{30.} W.H. Louisell, {\it Quantum Statistical Properties 
of Radiation\/} (Wiley, New York, 1973).

\item{31.} B.I. Shklovskii and A.L. Efros, {\it Electronic 
Properties of Doped Semiconductors\/} (Springer-Verlag, Berlin, 1984).

\item{32.} R. Kubo, M. Toda and N. Hashitsume, {\it Statistical 
Physics}, Vol. II (Springer-Verlag, Berlin, 1985).

\item{33.} A. Berg, M. Dobers, R.R. Gerhardts and K. von Klitzing,
Phys. Rev. Lett. {\bf 64}, 2563 (1990).

\item{34.} M. Dobers, K. von Klitzing, G. Weiman and K. Ploog,
Phys. Rev. Lett. {\bf 61}, 1650 (1988).

\item{35.} S.E. Barrett, G. Dabbagh, L.N. Pfeiffer, K.W. West and R. Tycko,
Phys. Rev. Lett. {\bf 74}, 5112 (1995).

}
\bye